\documentclass{jpsj3}
\usepackage{txfonts}
\usepackage{overcite}
\usepackage{amsmath} 
\usepackage{amssymb}
\usepackage{graphicx}
\usepackage{hyperref}
\usepackage{color}

\makeatletter
\providecommand\ext@figure{lof}\providecommand\ext@table{lot}
\let\ps@jpsj\ps@plain
\makeatother

\title{Fragment Size Density Estimator for Shrinkage-Induced Fracture Based on a Physics-Informed Neural Network}

\author{Shin-ichi Ito$^1$\thanks{ito@eri.u-tokyo.ac.jp}}
\inst{$^1$Earthquake Research Institute, The University of Tokyo, 1-1-1 Yayoi, Bunkyo-ku, Tokyo, Japan, 113-0032} 

\abst{
This paper presents a neural network (NN)-based solver for an integro-differential equation that models shrinkage-induced fragmentation.
The proposed method directly maps input parameters to the corresponding probability density function without numerically solving the governing equation, thereby significantly reducing computational costs.
Specifically, it enables efficient evaluation of the density function in Monte Carlo simulations while maintaining accuracy comparable to or even exceeding that of conventional finite difference schemes.
Validatation on synthetic data demonstrates both the method's computational efficiency and predictive reliability.
This study establishes a foundation for the data-driven inverse analysis of fragmentation and suggests the potential for extending the framework beyond pre-specified model structures.
}

\begin{document}
\maketitle

\section{Introduction}

Shrinkage-induced surface fractures in a thin surface layer are frequently encountered in both natural and industrial contexts.
They manifest in diverse forms, such as fissures in drying mud~\cite{Shorlin2000}, thermal cracks in glass~\cite{Yuse1993}, and cracks in paint coatings~\cite{Giorgiutti2016}.
These patterns are associated with a broad range of geological and material degradation processes~\cite{Goehring2015,Bacchin2018}.
The statistical characteristics of such patterns, particularly the fragment size distribution, offer crucial insights into the underlying mechanisms of pattern formation~\cite{Groisman1994,Allain1995,Lecocq2002,Akiba2017,Lilin2024}.
Analyzing these characteristics provides both fundamental understanding of fracture dynamics and practical value.
In particular, accurately estimating the fragment size distribution from limited experimentally accessible data remains a central challenge in inverse modeling and materials design.
Among these statistical features, the dynamical scaling law of fragment size distribution exemplifies the emergence of order within seemingly random fracture patterns~\cite{Ito2014a}.
As the fracturing proceeds, the mean fragment size decreases over time, and the fragment size distribution evolves accordingly.
Remarkably, when normalized by the instantaneous mean size, the distributions at different times collapse onto a single time-invariant curve, a hallmark of dynamical scaling.
This behavior suggests that universal features underlie shrinkage-induced fragmentation.
Although this phenomenon has been consistently observed in increasingly promising theoretical and numerical studies~\cite{Ito2014b,Halasz2017}, experimental confirmation remains preliminary~\cite{Ito2020}.
A theoretical framework capable of reproducing this scaling behavior was proposed by Ito and Yukawa~\cite{Ito2014b}.
Their stochastic model is formulated as an integro-differential equation (IDE) governing the probability density function of fragment sizes, explicitly incorporating the dynamics of individual fragment splitting.
By embedding a physics-motivated description of shrinkage-induced fracturing, the model provides a natural derivation of the dynamical scaling law, offering both theoretical consistency and physical plausibility.
However, obtaining accurate numerical solutions for this model is computationally demanding because of the high-dimensional integral terms involved.
Consequently, it becomes impractical to apply in scenarios requiring repeated model evaluations, such as Bayesian statistical inference based on Monte Carlo (MC) sampling.
This limitation underscores the need for more computationally efficient approximation strategies, particularly for the data-driven estimation of fragment size distributions and model parameters.

\begin{figure}[t]
 \centering
 \includegraphics[width=\linewidth]{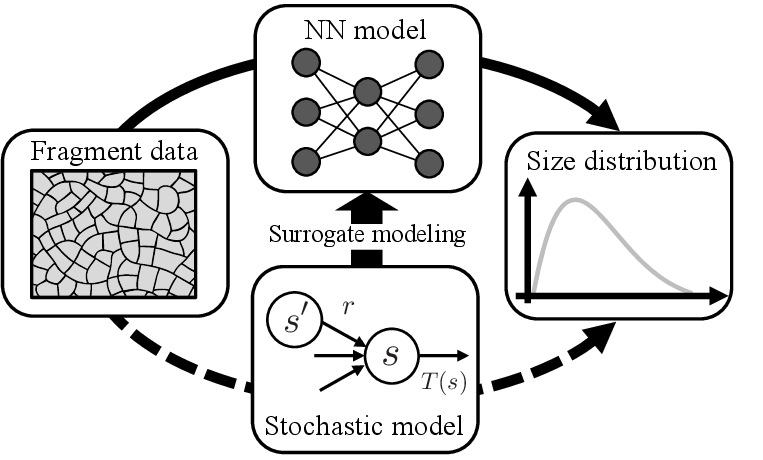}
 \caption{(Color online) Schematic illustration of the proposed method. The neural network (NN) model serves as a surrogate for the solution of a stochastic model describing the shrinkage-induced fragmentation process, thereby enabling the construction of the fragment size distribution.}
 \label{concept}
\end{figure}

In parallel with these developments, deep learning has witnessed rapid advances and has been increasingly adopted in the physical sciences, particularly for approximating complex models and accelerating computationally intensive simulations. In particular, neural networks (NNs) have proven effective as surrogate models that emulate the behavior of high-fidelity solvers with significantly reduced computational costs ~\cite{Raissi2017,Karniadakis2021,Kovachki2023}. Such approaches are particularly advantageous when repeated evaluations are required, such as in Bayesian inference and data assimilation.
Motivated by this, we propose an NN–based surrogate model that directly approximates the stochastic model (Fig.~\ref{concept}). Our approach is based on the framework of physics-informed NNs (PINNs)~\cite{Raissi2017}, which embed the governing equations into the training process. This enables the surrogate to be generalized across parameter ranges and eliminates the need for computationally expensive numerical integrations. Consequently, the model can be practically applied to inverse problems and probabilistic inferencing from experimental data, where repeated evaluations would otherwise be computationally prohibitive.

The remainder of this paper is structured as follows.  
Section~\ref{sec2} reviews the stochastic model proposed by Ito and Yukawa and summarizes its key properties.  
Section~\ref{sec3} introduces the proposed PINN-based surrogate model and details its training methodology and network architecture.  
Section~\ref{sec4} presents a quantitative evaluation of the surrogate, including comparisons with finite differences and exact solutions. It includes a demonstration of Bayesian inference using synthetic data.  
Section~\ref{sec5} concludes the paper with a summary of the findings and discusses future research directions.

\section{Model}\label{sec2}
This section briefly reviews the stochastic model for shrinkage-induced fracturing proposed by Ito and Yukawa~\cite{Ito2014b}. 
The model describes a stochastic process in which fragments are continuously divided into two pieces at a random ratio $r\in(0,1)$ over time.
The division ratio $r$ is assumed to be derived from the prescribed probability density $q(r)\in\mathbb{R}_{+}$.
Let $P(s;t)\in\mathbb{R}_{+}$ denote the probability density of fragments of size $s$ at time $t$.
The time evolution of $P(s;t)$ is governed by the balance between probability outflow and inflow owing to fragmentation:
\begin{equation}\label{eq2.1}
\frac{\partial P(s;t)}{\partial t} = -\frac{P(s;t)}{T(s)}+\int_{0}^{\infty}ds'\; w_{s'\rightarrow s} \frac{P(s';t)}{T(s')}.
\end{equation}
A schematic is shown in the bottom panel of Fig.~\ref{concept}.
Here, $T(s)$ is a size-dependent time scale representing the \emph{lifetime} of fragments of size $s$, i.e., the characteristic time over which such fragments disappear.
Kernel $w_{s' \rightarrow s}$ denotes the transition probability that a fragment of size $s'$ produces a smaller fragment of size $s$ via random division with ratio $r$, and is given by
\begin{equation}\label{eq2.2}
w_{s'\rightarrow s} = \int_{0}^{1}dr\; \delta(rs'-s)q(r),
\end{equation}
where $\delta(z)$ denotes the Dirac delta function.
Any density supported on $r\in(0,1)$ is admissible for $q(r)$.
Typically, a symmetric Beta distribution 
\begin{equation}\label{eq2.3}
q(r) = \frac{r^{\alpha-1}(1-r)^{\alpha-1}}{\text{Beta}(\alpha,\alpha)}
\end{equation}
is employed because of its computational convenience and symmetry with respect to $r=1/2$, where $\alpha$ is a positive parameter that controls the central concentration of probability density around $r=1/2$, and $\text{Beta}(v_{1},v_{2})=\int_{0}^{1}dw\; w^{v_{1}-1}(1-w)^{v_{2}-1}$.

Depending on the functional form of the lifetime $T(s)$, the solution of Eq.~\eqref{eq2.1} exhibits distinct behaviors that are physically relevant to fracturing~\cite{Ito2014b,Ito2015}.
When $T(s)$ is constant, the solution to \eqref{eq2.1} converges to a time-varying lognormal distribution.
In contrast, when $T(s) \propto s^{-\gamma}$ with $\gamma > 0$, the solution converges to a non-lognormal distribution that exhibits dynamical scaling:
\begin{equation}\label{dsl}
P(s;t)ds \xrightarrow[t\to\infty]{}  p\left(\frac{s}{\langle s \rangle_t}\right) \frac{ds}{\langle s \rangle_t},
\end{equation}
where $p(x)$ is a time-invariant scaling function and $\langle s \rangle_t = \int_{0}^{\infty}ds\; sP(s;t)$ denotes the mean size at time $t$.
These results are independent of the functional form of $q(r)$.
Such behavior is also observed in the shrinkage-induced fracturing of a thin elastic layer~\cite{Ito2014a,Halasz2017} and suggests that a power-law-decay lifetime underlies the dynamics of fragments.

This study has dual objectives: (i) to determine the functional form of the fragment size distribution when the dynamical scaling law holds, and (ii) to assess whether this functional form can be identified from actual fragment size data.
To analyze the asymptotic density under dynamical scaling, we consider Eq.~\eqref{eq2.1}, with $T(s) = t_{0}\,(s/s_{0})^{-\gamma}$, where $t_{0}$ and $s_{0}$ are characteristic time and size scales, respectively. 
We further assume that $q(r)$ follows Eq.~\eqref{eq2.3} and apply a scale transformation suited to a long time limit.
A straightforward scaling analysis shows that the mean fragment size behaves as $\langle s \rangle_{t} = (1/a)\, s_{0}\,(t/t_{0})^{-1/\gamma}$ over the given time limit, where $a$ is a constant dimensionless scale.
Appendix~\ref{app1} provides the derivation.
Let $x = s / \langle s \rangle_t$ denote mean-scaled fragment size.
By transforming the variable $s$ into $x$ in Eq.~\eqref{eq2.1} and dropping the time derivative in the long time limit, we obtain the following IDE for $p(x)$:
\begin{equation}\label{eq2.4.0}
I\left[x,p(x),a,\alpha,\gamma\right]=0,
\end{equation}
where 
\begin{equation}\label{eq2.4}
\begin{aligned}
&I\left[x,p(x),a,\alpha,\gamma\right]=\\
&-\frac{1}{\gamma}\frac{d}{dx}(xp(x)) - \left(\frac{x}{a}\right)^{\gamma}p(x)  
+ \int_{0}^{1}dr\,\frac{q(r)}{r^{1-\gamma}}\left(\frac{x}{a}\right)^{\gamma}p\left(\frac{x}{r}\right),
\end{aligned}
\end{equation}
subject to the normalization conditions for the density and its mean.
\begin{equation}\label{eq2.5}
\int_{0}^{\infty} dx\; p(x) = 1,
\end{equation}
\begin{equation}\label{eq2.6}
\int_{0}^{\infty} dx\; xp(x) = 1.
\end{equation}
Appendix~\ref{app2} provides the details of the derivation.
The objective is to determine the probability density $p(x)$ and the scale constant $a$ for a given parameter set $(\alpha, \gamma)$.
These equations are typically solved numerically by discretizing $x$ over a suitable support. However, several difficulties arise in practice.
The first lies in choosing an appropriate support.
The integral term in Eq.~\eqref{eq2.4} requires access to the values of $p(x)$ for $x \gg 1$ and the accuracy of $p(x)$ at $x \ll 1$ is heavily influenced by the integral.
Hence, a wide computational domain must be prepared to ensure stability.
However, this can incur a significant computational cost.
The second challenge is the need for careful discretization to maintain numerical stability.
The first and second terms in Eq.~\eqref{eq2.4} correspond to advection and damping terms in the original time-dependent equation.
Their discretization requires special care to avoid numerical instability even after dropping the time derivative.
Indeed, as will be demonstrated in Sect.~\ref{sec4.4}, insufficient care for certain parameter choices can yield spurious solutions that depart markedly from the true solution.
Although these difficulties can be addressed through trial and error, the procedure becomes particularly burdensome in settings, e.g., Bayesian inference based on MC simulations, where $p(x)$ must be evaluated repeatedly across a wide range of parameter sets.
To mitigate the reliance on trial-and-error approaches, we introduce a computational methodology based on physics-informed deep learning.

\section{Proposed method}\label{sec3}
Physics-informed deep learning is a mathematical framework that integrates physical models and NNs to solve governing equations efficiently and stably ~\cite{Raissi2017,Kovachki2023}.
A representative realization of this framework is the PINN~\cite{Raissi2017}, which embeds the structure of partial differential equations or IDEs directly into the loss function of an NN.
This enables the network to learn solutions that are consistent with the underlying physical laws without relying on supervised data.

We adopted the PINN approach to solve the IDE (Eq.~\eqref{eq2.4.0}) that governs the scaled fragment size distribution.
For this, NNs were constructed that approximate both the solution $p(x)$ and scale constant $a$, and trained them by minimizing the loss function that enforces Eq.~\eqref{eq2.4} along with the normalization conditions given in Eqs.~\eqref{eq2.5} and~\eqref{eq2.6}.
This method eliminates the need for manual discretization and support tuning, offering a flexible and efficient alternative to conventional numerical schemes.
The mathematical formulation of the proposed PINN-based method, the loss-function design, and the treatment of the integro-differential operator are detailed below.

Let $p_{\theta}(x;\alpha,\gamma)$ and $a_{\phi}(\alpha,\gamma)$ denote the NNs parameterized by $\theta$ and $\phi$, representing the solution components.
These networks are trained to satisfy Eqs.~\eqref{eq2.4.0}--\eqref{eq2.6} by minimizing the following loss terms:
\begin{equation}\label{l1}
l_{1}(\theta,\phi;\alpha,\gamma) = \frac{1}{2}\int_{0}^{\infty} dx\; g(x)\left(I\left[x, p_{\theta}(x;\alpha,\gamma), a_{\phi}(\alpha,\gamma), \alpha, \gamma\right]\right)^{2},
\end{equation}
\begin{equation}\label{l2}
l_{2}(\theta;\alpha,\gamma) = \frac{1}{2}\left(1 - \int_{0}^{\infty} dx\; p_{\theta}(x;\alpha,\gamma) \right)^2,
\end{equation}
\begin{equation}\label{l3}
l_{3}(\theta;\alpha,\gamma) = \frac{1}{2}\left(1 - \int_{0}^{\infty} dx\; x p_{\theta}(x;\alpha,\gamma) \right)^2.
\end{equation}
Here, $l_1$ represents the residual of the IDE (Eq.~\eqref{eq2.4.0}), whereas $l_2$ and $l_3$ impose the normalization constraints (Eqs.~\eqref{eq2.5} and~\eqref{eq2.6}), respectively.
Weight $g(x)$ can be any positive function. This study employs $g(x) = 1/x$ to facilitate learning in regions where $x$ is small. 
These weightings have been empirically confirmed to prevent overtraining.
The training across a range of parameter sets $(\alpha, \gamma)$ is generalized by defining the total loss $L$ as the expectation over log-uniform distributions:
\begin{equation}\label{losspiece}
l(\theta,\phi;\alpha,\gamma) =  l_1(\theta,\phi;\alpha,\gamma) + w_2 l_2(\theta;\alpha,\gamma) + w_3 l_3(\theta;\alpha,\gamma), 
\end{equation}
\begin{equation}\label{loss}
L(\theta,\phi) = \underset{\substack{\alpha\sim \text{LU}[\alpha_{\min},\alpha_{\max}],\\
\gamma\sim \text{LU}[\gamma_{\min},\gamma_{\max}]}}{\mathbb{E}}\left[ l(\theta,\phi;\alpha,\gamma) \right],
\end{equation}
where $\text{LU}[u_{\min}, u_{\max}]$ denotes the log-uniform distribution supported on the interval $[u_{\min}, u_{\max}]$ and $w_2$ and $w_3$ are positive scalar weights, which are set to $w_2=1$ and $w_3=1$ in this study.
This log-uniform-based expectation has been empirically demonstrated to facilitate efficient training over a wide parameter range and has been adopted in several studies related to generative tasks~\cite{Dosovitskiy2019,Bae2022}.
The optimization of $L$ with respect to $\theta$ and $\phi$ yields an NN surrogate model capable of predicting $p(x)$ across a broad range of parameters, without explicitly solving Eq.~\eqref{eq2.4.0}.

The NN architecture is crucial for achieving both robust and rapid convergence.
Let $\text{NN}_{p}(x,\alpha,\gamma;\theta)$ and $\text{NN}_{a}(\alpha,\gamma;\phi)$ be base NNs without architectural constraints.
Thus, we define the representations of $p_{\theta}(x;\alpha,\gamma)$ and $a_{\phi}(\alpha,\gamma)$ as follows:
\begin{equation}\label{2.7}
\begin{aligned}
p_{\theta}(x;\alpha,\gamma) &= \exp\left[ (\alpha - 1) \text{NN}_{p}(x,\alpha,\gamma;\theta) \right] \hat{p}_{\alpha,\gamma}(x), \\
a_{\phi}(\alpha,\gamma) &= \exp\left[ (\alpha - 1) \text{NN}_{a}(\alpha,\gamma;\phi) \right] c_{\alpha,\gamma},
\end{aligned}
\end{equation}
where $\hat{p}_{\alpha,\gamma}(x)$ is a generalized gamma distribution and $c_{\alpha,\gamma}$ is its scale constant, defined as follows:
\begin{equation}\label{2.8}
\begin{aligned}
\hat{p}_{\alpha,\gamma}(x) &= \frac{\gamma}{c_{\alpha,\gamma}\Gamma(\alpha/\gamma)} \left( \frac{x}{c_{\alpha,\gamma}} \right)^{\alpha - 1} \exp\left[ - \left( \frac{x}{c_{\alpha,\gamma}} \right)^{\gamma} \right], \\
c_{\alpha,\gamma} &= \frac{\Gamma(\alpha/\gamma)}{\Gamma((\alpha+1)/\gamma)},
\end{aligned}
\end{equation}
$\Gamma(z)$ denotes the gamma function.
The probability density $\hat{p}_{\alpha,\gamma}$ serves as an approximate solution to Eq.~\eqref{eq2.4.0}, which captures the asymptotic behaviors as $x \to 0$ and $x \to \infty$.
Appendix~\ref{app3} details the derivation of the asymptotic behaviors.
Notably, this coincides with the exact solution when $\alpha = 1$.
Embedding the structure of Eq.~\eqref{2.8} into the network representations, as shown in Eq.~\eqref{2.7}, significantly accelerates training convergence compared with optimizing a vanilla NN without prior knowledge of the solution.
Moreover, the rapid decay induced by the stretched exponential term in Eq.~\eqref{2.8} suppresses the growth of $p_{\theta}$ as $x \to \infty$, which may otherwise hinder the numerical evaluation of integrals in the loss function, thereby contributing to stable optimization.

\section{Results and Discussion}\label{sec4}

\subsection{Setting for numerical experiments}\label{sec4.1}

\begin{figure}[t]
 \centering
 \includegraphics[width=\linewidth]{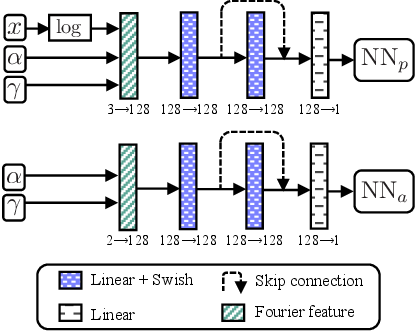}
 \caption{(Color online) Schematic architecture of the neural networks used in this study.}
 \label{architecture}
\end{figure}

Figure~\ref{architecture} illustrates the neural network architecture used in this study.
The neural networks $\text{NN}_{p}(x,\alpha,\gamma;\theta)$ and $\text{NN}_{a}(\alpha,\gamma;\phi)$ are designed as the combination of a random Fourier feature (RFF) encoder~\cite{Tancik2020} for the initial layer and multilayer perceptron (MLP) with a skip connection.
The $\text{NN}_p$ design internally transforms the input variable $x$ into $\log(x)$ before it is passed to the encoder.
The RFF encoder contains 128 $\sin$ and $\cos$ features with random frequencies drawn from a standard normal distribution. 
These encoded features are then processed by a three-layer MLP in which every linear transformation is followed by a Swish activation~\cite{Ramachandran2017}: an input block Linear ($128\rightarrow 128$) $+$ Swish; a second block Linear ($128\rightarrow 128$) $+$ Swish augmented with a skip connection from the first block; and a final linear projection Linear ($128\rightarrow 1$) that maps to the output space.
The weights and biases in each Linear are initialized using small random values sampled from a normal distribution with a mean of zero and variance $10^{-6}$.
Weight scaling~\cite{Wang2022} is introduced to each Linear to facilitate learning. 
This architecture was selected because it achieved the best performance across our experiments, which varied both the layer width and the number of skip-connected layers.
A comprehensive analysis of this architecture-dependency study is provided in Section 1 of the Supplemental Materials~\cite{Supplemental}.

To ensure numerical accuracy, the integrals involved in Eq.~\eqref{eq2.4} and Eqs.~\eqref{l1}--\eqref{l3} are evaluated using double exponential formulas~\cite{Takahasi1974}.
For the integrals in the losses~\eqref{l1}--\eqref{l3}, which are defined over $x \in (0, \infty)$, we apply the variable transformation $x = \exp(\pi\sinh(y)/2)$ with $y \in \mathbb{R}$.
Similarly, the integral appearing in $I$ (Eq.~\eqref{eq2.4}), defined over $r \in (0, 1)$, is computed using the transformation $r = \left[1+ \tanh(\pi\sinh(z) / 2)\right]/2$ with $z \in \mathbb{R}$.
The transformed coordinates $y$ and $z$ are uniformly discretized over the intervals $y \in [-3, 2]$ and $z \in [-3, 3]$, respectively.
The integral in Eq~\eqref{l1} is discretized into $64$ segments and evaluated by simply summing the discretized values. 
The integrals in Eqs~\eqref{eq2.4},~\eqref{l2}, and~\eqref{l3} are discretized into $32$ segments.
Within each segment, 16-point Gauss–Legendre (GL) quadrature is applied.
The integral in Eq.~\eqref{l1} is discretized more sparsely than the others because the neural network's interpolation capability makes a dense set of quadrature points unnecessary, which in turn lowers the subsequent optimization cost.
The expectation in Eq.~\eqref{loss} is approximated by discretizing the log-scale domains $\alpha\in[1.2,6.0]$ and $\gamma\in[1.2,6.0]$ over $15\times 15$ cells, which means that $16\times 16$ grid points are used. 
In each cell, the loss $l$ (Eq~\eqref{losspiece}) is evaluated at the vertices and interpolated as a bilinear function. 
Using the bilinear interpolated loss, the expectation is computed using a 5-point GL quadrature for each dimension.
All computations associated with the PINN-based model are performed with double precision using Python 3.9, CUDA 12.6, and PyTorch 2.6 on an NVIDIA H200 SXM GPU.
For the sake of reproducibility and to support future work by other researchers, the main code has been made available online~\cite{Itogithub2025}.

\subsection{Optimization of networks}\label{sec4.2}

This section presents and analyzes the results of optimizing the PINN-based model.
The choice of optimizer is a key determinant of both accuracy and computational efficiency.
Among several options, the limited-memory Broyden--Fletcher--Goldfarb--Shanno (L-BFGS) algorithm~\cite{Nocedal1980} has been widely adopted in PINN research~\cite{Krishnapriyan2021,lu2021}.
Being a quasi-Newton method, it approximates the inverse Hessian using a set of auxiliary vectors accumulated during optimization.
The ability to achieve near-quadratic convergence using only gradient information underlies its appeal.
However, for certain problems, convergence may stagnate in the later stages because of inaccurate Hessian approximations, thereby limiting its effectiveness.
An alternative Newton-like method that also relies solely on gradient information is the Levenberg–-Marquardt (LM) algorithm~\cite{LEVENBERG1944,Marquardt1963}.
Designed to minimize the loss functions expressed as the sum of squared residuals (SSR), the LM algorithm achieves quadratic convergence near the optimum.
As many PINN problems, including those considered here, can be formulated as SSR minimization, LM is an ideal candidate.
However, the standard LM method requires solving a large linear system at each iteration, involving a Gramian matrix, the size of which scales with the number of network parameters.
This makes its direct application infeasible for large-scale NNs, owing to excessive memory and computational demands.
To address this limitation, a recently proposed technique~\cite{Jnini2025} reformulates the problem in the residual space rather than in the parameter space, enabling large-scale LM optimization without approximation.
This development renders LM-based methods practically applicable to PINN problems.
The details of the algorithm are shown in Section 2 of the Supplemental Materials~\cite{Supplemental}.
To obtain a high-performing model, this study adopted a hybrid approach combining L-BFGS and LM based on this technique.
The model's generalization capability during optimization was evaluated by computing the following test loss function
\begin{equation}\label{testloss}
L^{\text{test}}_{\alpha,\gamma} = \int_{0}^{\infty}dx\, \left[p_{\theta}(x;\alpha,\gamma)-p_{\alpha,\gamma}(x)\right]^{2},
\end{equation}
for the parameter pairs $(\alpha,\gamma)=(2,1)$ and $(2,2)$, representing the cases in which the exact analytical solutions to Eq.~\eqref{eq2.4.0} are available.
The function $p_{\alpha,\gamma}(x)$ denotes the exact solution of Eq.~\eqref{eq2.4.0}, which is explicitly expressed by
\begin{equation}\label{sol21}
p_{\alpha,\gamma}(x)=\frac{432}{25}x\left[E_{2}\left(\frac{6}{5}x\right)-E_{3}\left(\frac{6}{5}x\right)\right], 
\end{equation}
for $(\alpha,\gamma)=(2,1)$ and 
\begin{equation}\label{sol22}
p_{\alpha,\gamma}(x)=\frac{27\pi}{32}x\,\exp\left(-\frac{9\pi}{64}x^2\right)-\frac{81\pi^{2}}{256}\,x^2\,\text{erfc}\left(\frac{3\sqrt{\pi}}{8}x\right),
\end{equation}
for $(\alpha,\gamma)=(2,2)$, where $E_{n}(x) = \int_{1}^{\infty}dw\, w^{-n}e^{-wx}$ for $n\in\mathbb{N}$ and $\text{erfc}$ is the complementary error function.
Note that these parameter pairs are not used in computing the discretized $L$, i.e., evaluating $L_{2,1}^{\text{test}}$ ($L_{2,2}^{\text{test}}$) corresponds to investigating the extrapolation (interpolation) ability.
The learning domain $[1.2,6.0]^{2}$ defined in Sect.~\ref{sec4.1} was chosen to quantify these abilities.

\begin{figure}[t]
 \centering
 \includegraphics[width=0.65\linewidth]{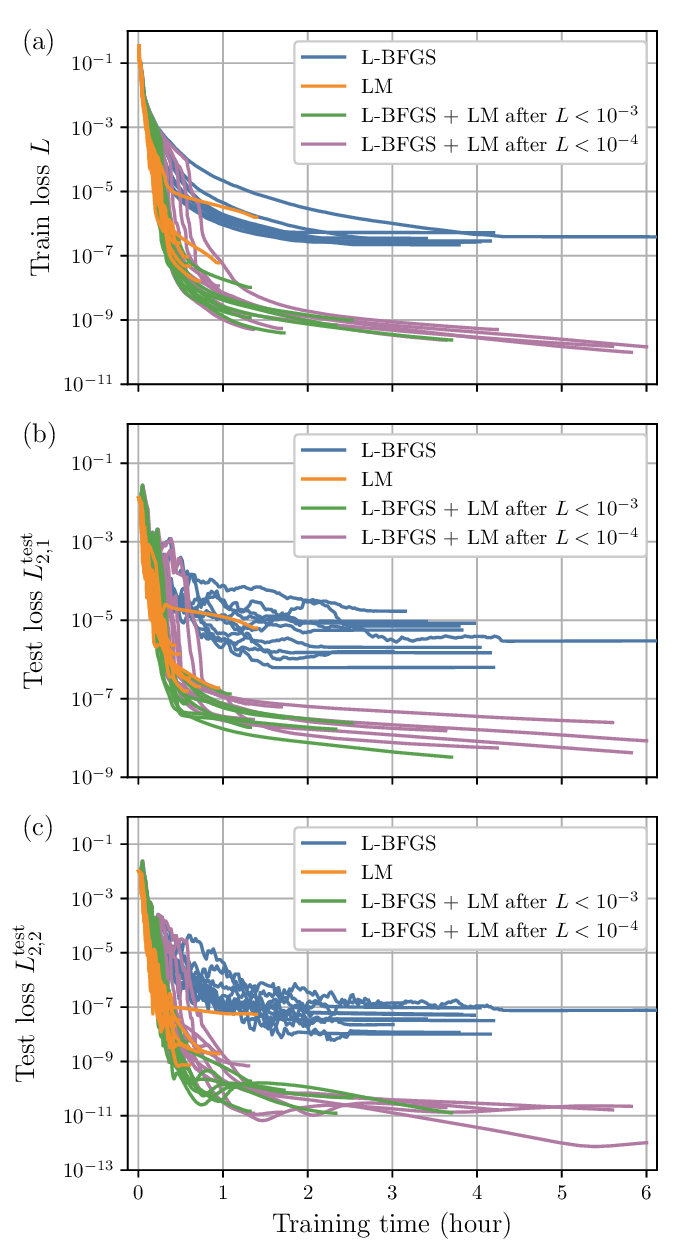}
 %
 \caption{
 (Color online) Optimization history of each loss as a function of actual training time. Panels (a), (b), and (c) show the train loss $L$, the test loss $L^{\text{test}}_{2,1}$, and $L^{\text{test}}_{2,2}$, respectively. The line colors indicate different optimization strategies. Each strategy underwent ten trials with different random seeds to initialize the network parameters.
 }
 \label{losshistory}
\end{figure}

Figure~\ref{losshistory} shows the optimization history of each loss as a function of the actual training time.
This experiment compared four optimization strategies:
(i) L-BFGS only;
(ii) LM only;
(iii) L-BFGS followed by LM refinement once the training loss $L$ falls below a tolerance of $10^{-3}$; and
(iv) identical to (iii) but with a stricter tolerance of $10^{-4}$.
Each strategy underwent ten trials using different random seeds to initialize the network parameters.
In applying strategy (i), the loss decreased smoothly; however, the rate of decrease diminished as the optimization progressed.
In contrast, strategy (ii) showed a faster initial reduction than strategy (i), but was often trapped in suboptimal local minima, resulting in premature convergence with relatively high final loss values.
Although some variation in the final loss values was observed in hybrid strategies (iii) and (iv), both strategies effectively overcame the stagnation encountered in strategy (i) and achieved faster convergence to drastically smaller values.
No substantial performance difference was noted between (iii) and (iv); instead, the final outcomes appeared to depend more strongly on the random initialization of the network parameters.
A similar trend was observed for the test loss; the final test loss values obtained by strategies (iii) and (iv) were consistently lower than those of strategies (i) and (ii), achieving improvements of one to five orders of magnitude.

\subsection{Comparison with finite difference scheme}\label{sec4.3}

This section compares the solutions obtained by the best-performing PINN model with those obtained by a conventional non-deep-learning-based method in terms of accuracy and prediction time.
From the previous section, we selected the model with the smallest final train loss as the best-performing PINN model.
As a baseline for comparison, we solve Eq.~\eqref{eq2.4.0} for the given parameter set of $(\alpha,\gamma)$ using a finite difference (FD) scheme.
To obtain this solution, we reinterpret Eq.~\eqref{eq2.4.0} as the convergence of a time-dependent problem by introducing a pseudo-time derivative, i.e., $\partial p/\partial \hat{t}=I$, where $\hat{t}$ is the pseudo-time.
Since Eq.~\eqref{eq2.4}, which includes the $r$-integral, is difficult to construct an appropriate discretization, Eq.~\eqref{eq2.4} is transformed into an equivalent form:
\begin{equation}\label{eq2.4.a}
\begin{aligned}
&I\left[x,p(x),a,\alpha,\gamma\right]=\\
&-\frac{1}{\gamma}\frac{d}{dx}(xp(x)) - \left(\frac{x}{a}\right)^{\gamma}p(x)  
+ \int_{x}^{\infty}\frac{dw}{w}\,q\left(\frac{x}{w}\right)\left(\frac{w}{a}\right)^{\gamma}p\left(w\right),
\end{aligned}
\end{equation}
which only requires discretization in the $x$-direction.
The pseudo-time-dependent IDE using this $I$ is integrated until convergence starting from the initial guess $p(x)=e^{-x}$.
Similar to that mentioned in Sect.~\ref{sec4.1}, the variable $x$ is transformed using the double exponential formula, and then uniformly discretized into $k$ grid points.
In this experiment, the number of grid points $k$ is a control parameter.
The $x$-derivative term that appears as the first term on the right-hand side of Eq.~\eqref{eq2.4.a} is discretized using a first-order upwind scheme, assuming a zero-curvature condition at the left boundary of the $x$-domain.
Owing to the extreme stiffness of the problem, we employed a first-order exponential integrator with adaptive step-size control for time integration.
The solution is rescaled at every time step to satisfy the normalization constraint (Eq.~\eqref{eq2.5}).
The value of the scale constant $a$, which depends on the parameter set, is unknown in advance and cannot be directly determined from the mean normalization condition (Eq.~\eqref{eq2.6}).
In practical computation, this pseudo-time-dependent problem is solved by assuming $a=1$, and then rescaling the obtained solution to satisfy Eq.~\eqref{eq2.6}.
All computations were performed in the environment described in Sect.~\ref{sec4.1}.
The repeated matrix computations in this scheme were GPU-accelerated using CuPy 13.3.

\begin{figure}[t]
 \centering
 \includegraphics[width=\linewidth]{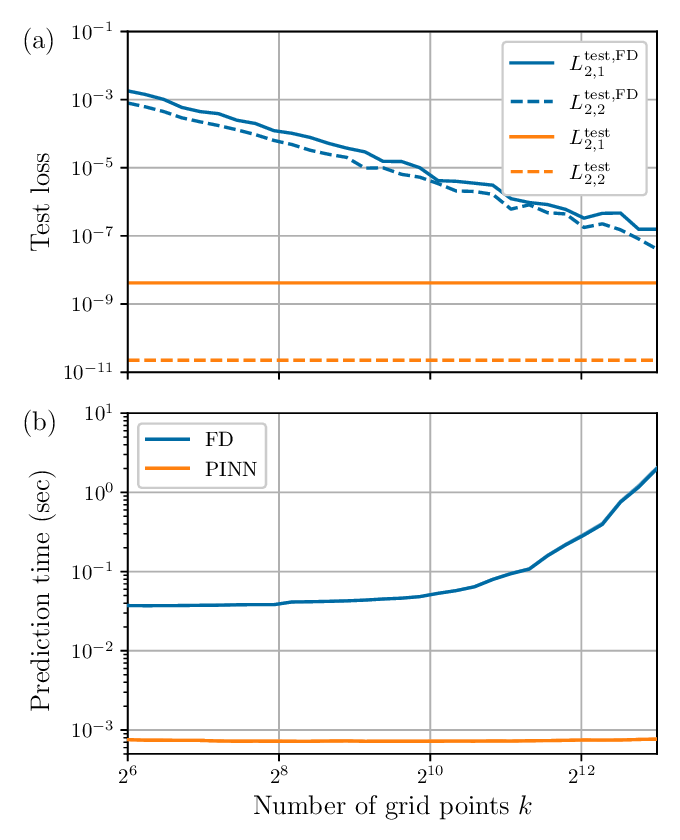}
 \caption{(Color online) Comparison with the finite difference (FD) scheme.
Panel (a) shows the test losses of the best-performing PINN model obtained from the analysis in Sect.~\ref{sec4.2} and those of the FD model as functions of the number of grid points.
Panel (b) presents the prediction time required to obtain solutions using each model.
Each line in (b) indicates the 50th percentile over 1,000 parameter sets uniformly sampled from $[1.2, 6.0]^{2}$.
The shaded regions in (b) represent the 5th to 95th percentile intervals, although they may not be visually distinguishable due to their narrow width.}
 \label{compare}
\end{figure}

Analogous to Eq.~\eqref{testloss}, the test loss for the FD solution is defined as
\begin{equation}\label{testlossFD}
L^{\text{test,FD}}_{\alpha,\gamma}(k) = \int_{0}^{\infty}dx\, \left[p^{\text{FD},k}_{\alpha,\gamma}(x)-p_{\alpha,\gamma}(x)\right]^{2},
\end{equation}
for $(\alpha,\gamma)=(2,1)$ and $(2,2)$, where $p^{\text{FD},k}_{\alpha,\gamma}(x)$ denotes the piecewise linear interpolation of the FD solution corresponding to a given set of $k$ grid points.

Figure~\ref{compare}(a) shows the test losses $L^{\text{test,FD}}_{2,1}$ and $L^{\text{test,FD}}_{2,2}$ as functions of the number of grid points $k$, alongside the corresponding test losses $L^{\text{test}}_{2,1}$ and $L^{\text{test}}_{2,2}$ of the best-performing PINN model.
The FD test losses exhibit a power-law decay with respect to $k$; however, achieving the accuracy level of the PINN model requires substantially more grid points.
Notably, at $k = 2^{6}$, which corresponds to the resolution used for the IDE during PINN training, the PINN test losses were four to eight orders of magnitude smaller than those of the FD method, reflecting the superior interpolation capability of NNs.

Figure~\ref{compare}(b) shows the prediction time required to obtain solutions using each model.
The results summarize the statistics from 1,000 trials using parameter sets uniformly sampled from the domain $[1.2, 6.0]^2$, which is consistent with the training domain.
To ensure a fair comparison, the FD and PINN models were both evaluated over the same grid size range $k$.
Both prediction time curves have extremely narrow error bands, indicating that both methods are robust to changes in the parameter sets.
For the FD model, the prediction time remains nearly constant for a small $k$ but increases rapidly beyond $k \approx 2^{10}$, primarily because of the computational cost of evaluating the matrix exponential in the exponential integrator.
In contrast, the prediction time of the PINN model remains nearly constant across the entire range considered and is approximately $1/50$ of the minimum FD prediction time.  
Theoretically, the prediction time of the PINN model scales linearly with $k$; however, this proportionality is not apparent in the range of $k$ considered here because the prediction time is dominated by fixed overhead.
Moreover, reaching a test loss comparable to that of the PINN model requires at least $k > 2^{13}$ in the FD scheme, which results in a prediction time that is more than three orders of magnitude longer.  
These results clearly demonstrate the advantages of the PINN model in terms of both prediction speed and accuracy.
As discussed later, the training cost is not a limiting factor in scenarios such as MC sampling, where a large number of forward predictions are required.

\subsection{Solution profiles and errors}\label{sec4.4}

This section examines the IDE solution profiles across several representative parameter sets, including those without exact solutions.
To assess the accuracy of the PINN beyond the aggregate test losses, we compare its predictions with those of the FD method, and where available, with exact solutions.

\begin{figure*}[t]
 \centering
 \includegraphics[width=\textwidth]{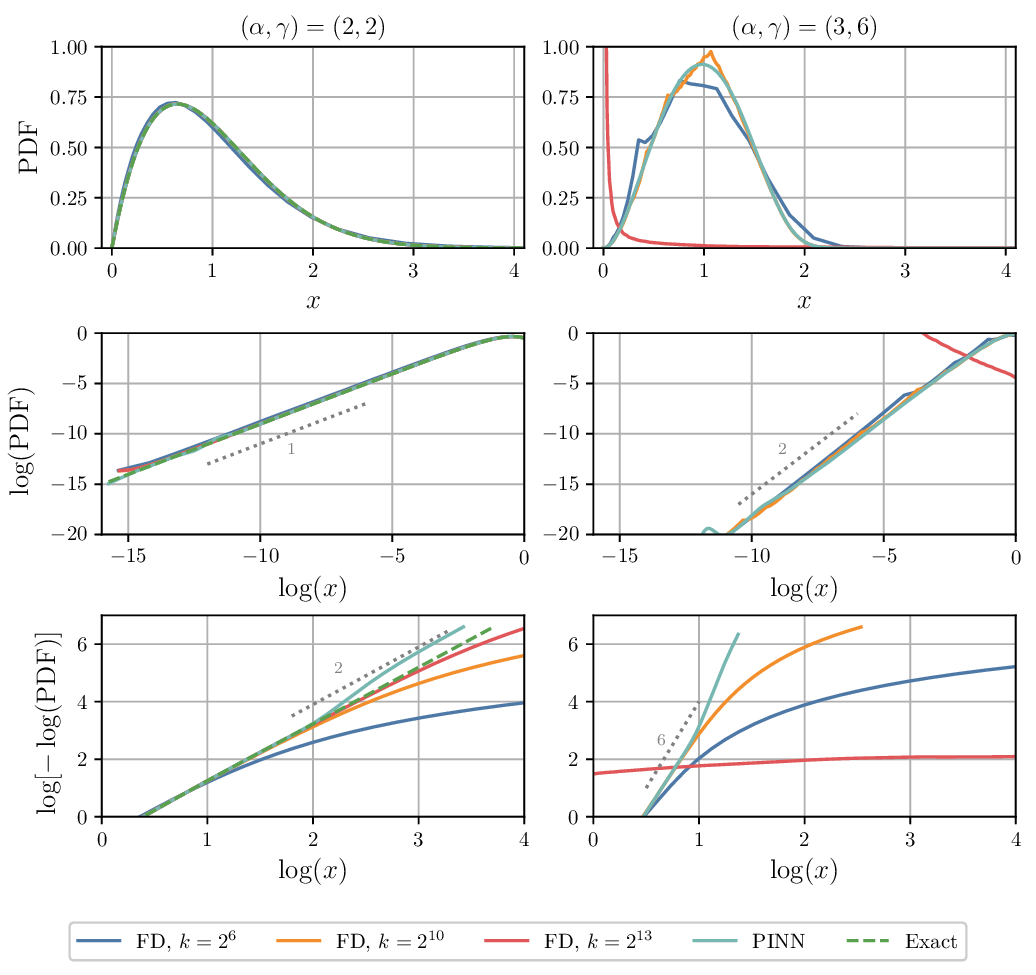}
 \caption{(Color online) Detailed solution profiles.
The first and second columns correspond to $(\alpha,\gamma) = (2,2)$ and $(3,6)$, respectively.
Each row displays the profiles of the probability density function (PDF) on different scales: linear (top), $\log$–$\log$ (middle), and $\log$–$\log(-\log)$ (bottom), where the bottom row is suited for capturing stretched exponential behavior.
In the second row, the dotted lines indicate the power-law exponents; in the third row, they represent the stretched exponential exponents.
The theoretical asymptotic behaviors are given by $p(x)\propto x^{\alpha-1}$ as $x\rightarrow 0$ and $p(x)\propto \exp\left[-(x/a)^{\gamma}\right]$ as $x\rightarrow \infty$.
}
 \label{solutions}
\end{figure*}

Figure~\ref{solutions} illustrates the solution profiles obtained by each method for $(\alpha,\gamma) = (2,2)$ and $(3,6)$, plotted on various scales to highlight different aspects of the solution behavior.
The first column corresponds to the case $(\alpha,\gamma) = (2,2)$.
As the number of grid points increases, the FD solutions progressively approach the exact solution.
The PINN solution also demonstrates excellent agreement with the exact solution, except in regions where the values are extremely small.
Notably, the asymptotic behaviors of $x \to 0$ and $x \to \infty$ are well captured.
In contrast, the second column presents the case $(\alpha,\gamma) = (3,6)$.
Here, the FD solutions exhibit numerical instability, particularly at $k = 2^{13}$, where they converge to a spurious solution that deviates markedly from the expected asymptotic behavior, probably because of the stiffness in the equation induced by the large value of $\gamma$.
By comparison, the PINN solution successfully captures the anticipated asymptotic structure.
Although the exact solution is not available for this parameter set, the ability of the PINN to reproduce the expected asymptotic behavior suggests that it closely approximates the true solution.
This contrast may be attributed to PINNs optimizing a global function representation rather than solving a discretized system, which can mitigate numerical instabilities caused by stiffness.
These results highlight the strong potential of PINNs to solve the IDE robustly across a wide range of parameter values owing to the high approximation capability of NNs.
Such robustness is particularly valuable in applications that require repeated solution evaluations over varying parameters, such as Bayesian inference based on MC sampling.

The appearance of spurious solutions in the FD scheme highlights the value of physics-informed learning, which embeds physical equations as constraints. A neural-network regression trained solely on FD-generated data, without such constraints, can readily absorb these spurious patterns, producing outputs that fit the data yet violate the governing equation. In addition, because such a regression model achieves only local fits, it generally requires densely sampled data to avoid overtraining. By incorporating the governing equation, the PINN formulation enables the model to embed global physical structure, suppress overtraining, and yield accurate predictions.

\subsection{Synthetic Monte Carlo benchmark}\label{sec4.5}

\begin{figure}[t]
 \centering
 \includegraphics[width=\linewidth]{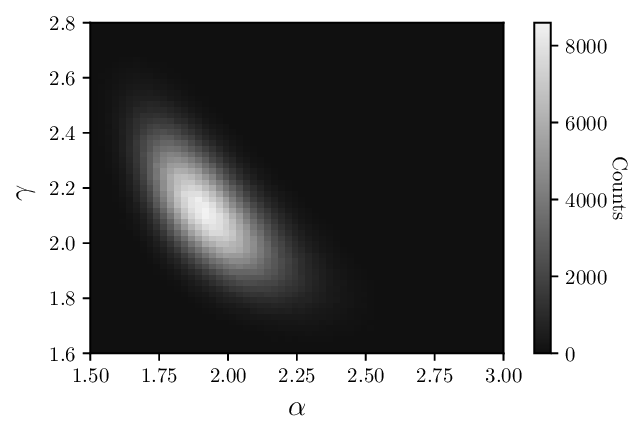}
 \caption{(Color online) Histogram view of the joint posterior distribution of $\alpha$ and $\gamma$ obtained from an MC simulation applied to 1,000 synthetic $x$-samples drawn from the exact density for $(\alpha,\gamma)=(2,2)$ (Eq.~\eqref{sol22}).
 The pixels in this color map correspond to the histogram bins.
 }
 \label{posterior}
\end{figure}

The speed and robustness demonstrated in Sects.~\ref{sec4.3} and~\ref{sec4.4} render the PINN-based approach particularly suitable for Bayesian inference via MC sampling.
To illustrate its practical utility, we evaluated the PINN-based Bayesian estimator's performance on a synthetic dataset comprising $n = 1,000$ samples, $\mathcal{D} = \{x_1, x_2, \dots, x_n\}$, drawn from the exact density $p_{2,2}(x)$ (Eq.~\eqref{sol22}).
The joint posterior distribution $\rho(\alpha, \gamma \mid \mathcal{D})$ is given by
\begin{equation}
\rho(\alpha, \gamma \mid \mathcal{D}) \propto R(\alpha, \gamma) \prod_{i=1}^{n} p_{\theta}(x_i; \alpha, \gamma),
\end{equation}
where $R(\alpha, \gamma)$ denotes the prior distribution, which is assumed to be uniform over the same ranges used during training.
We employed the Metropolis--Hastings algorithm with ten million proposals, thinned by retaining every fifth state, resulting in two million posterior samples.
Each parameter was updated using a normal proposal distribution with mean zero and variance $0.02^{2}$.
Figure~\ref{posterior} presents a histogram of the posterior samples, offering a high-resolution view enabled by the large effective sample size.
The full inference run was completed within approximately 2 hours.
By contrast, performing the same inference using an FD solver for likelihood evaluation requires $\ge$ 100 hours on the same hardware, even when using a coarse grid with limited accuracy.
If a higher-resolution FD solver is employed to match the accuracy of the PINN, the estimated runtime exceeds 5,000 hours.
In comparison, as shown in Sect.~\ref{sec4.2}, training the PINN requires at most around 6 hours, which is negligible relative to the total cost of posterior sampling.
Moreover, FD-based solvers are susceptible to numerical instability, particularly in stiff regimes, which can produce spurious solutions.
Such inaccuracies may destabilize the MC procedure or introduce bias into posterior estimates.
In contrast, the PINN-based model offers robust and accurate performance across a wide range of parameters, supporting stable and reliable MC inferences.
This combination of computational efficiency and numerical robustness enables high-fidelity Bayesian estimation at a fraction of the computational cost.
Although this benchmark considers a single MC run, the proposed method is expected to accelerate applications that require repeated inferences, such as Bayesian free-energy estimation, where many independent simulations are necessary.
Furthermore, because the PINN-based likelihood is differentiable with respect to $\alpha$ and $\gamma$, the Metropolis--Hastings algorithm can be replaced with gradient-based samplers, such as the Hamiltonian MC~\cite{Betancourt2017} or the zigzag sampler~\cite{Corbella2022}, to accelerate the inference further.
Such enhancements are not feasible using FD-based solvers.

\section{Conclusions}\label{sec5}
We proposed a PINN–based solver for an IDE that characterizes shrinkage-induced fragmentation.
The proposed framework substantially accelerates inference while maintaining accuracy comparable to using conventional finite-difference methods.
By directly mapping the input parameters to the associated probability density function without explicitly solving the governing equation, the model enables a significant reduction in computational cost, particularly for MC simulations that require repeated density evaluations.
Although the present study focused on methodological development and validation using synthetic data, a natural direction for future research is to apply the proposed framework to inverse problems involving experimentally observed fragment-size distributions.
In this study, the MC tests deliberately used noise-free pseudo-data. In contrast, real measurements always contain stochastic errors.
Such uncertainty must be addressed at the statistical inference layer---for example, by adopting a likelihood that accommodates Poisson or Gaussian noise---rather than at the forward-solver layer itself.
As the proposed PINN only replaces the deterministic kernel inside that likelihood, its speed advantage and accuracy remain intact once an appropriate noise-tolerant Bayesian model is in place.

Another direction for future research is to extend the current framework to solve the original time-dependent integral equation directly (Eq.~\eqref{eq2.1}), thereby enabling simultaneous resolution of both the parameter and temporal dimensions.
Although this approach is currently computationally demanding during training owing to the high-dimensional integration required in the loss function, such a model would enable the identification of transitions toward dynamical scaling directly from time-series fragment data.

In addition, a promising methodological extension lies in relaxing the assumption of the prescribed functional forms for $q(r)$ and $T(s)$.
By incorporating recent advances in operator learning~\cite{Kovachki2023}, it may be possible to construct a generalized framework that accepts $q(r)$ and $T(s)$ as functional inputs and predicts the corresponding probability density.
This extension would enable the discovery of hidden physical processes directly from the data, thereby enhancing the expressiveness and applicability of the model beyond the constraints of pre-defined constitutive assumptions.

\section*{Acknowledgments}
We used ABCI 3.0, provided by AIST for the GPU computations. This work was supported by JSPS KAKENHI (Nos. JP22K03542, JP23H00466, and JP24K02951); ERI JURP (Nos. 2025-B-01, 2024-B-01, and 2025-A-03); the MEXT Project for Seismology Toward Research Innovation with Data of Earthquake (STAR-E) (No. JPJ010217); and JST NEXUS (No. JPMJNX25C5).

\appendix

\section{Time evolution of mean fragment size}\label{app1}
In this appendix, we outline the derivation of the functional form governing the time evolution of the mean fragment size, as presented in Ito and Yukawa~\cite{Ito2014b}.
The derivation begins by imposing a scale-invariant condition on Eq.~\eqref{eq2.1}.
Consider a scale transformation in the form $P(s;t) \rightarrow P(\beta s; \eta t)$, where $\beta$ and $\eta$ are time-dependent scaling factors.
To preserve the form of Eq.~\eqref{eq2.1} under this transformation, the lifetime function $T(s)$ must satisfy the condition
\begin{equation}
\frac{\eta}{T(\beta s)} = \frac{1}{T(s)}.
\end{equation}
Solving this condition yields $T(s) = T(1)\,s^{-\gamma}$, where $\gamma \coloneqq -T'(1)/T(1)$, and the time scaling factor for time is given by $\eta = \beta^{-\gamma}$.
Assuming $\eta = 1/t$, we obtain
\begin{equation}\label{a2}
P(s;t)ds = \beta P(\beta s; \eta t)ds = t^{1/\gamma}P(t^{1/\gamma} s; 1)ds.
\end{equation}
The mean fragment size $\langle s \rangle_t = \int_{0}^{\infty} ds\, s\,P(s;t)$ is evaluated by substituting Eq.~\eqref{a2} into its definition, yielding
\begin{equation}\label{a3}
\langle s \rangle_t = \mathcal{F}(t) = \beta \int_{0}^{\infty} ds\; sP(\beta s; \eta t).
\end{equation}
By changing the variable $z = \beta s$ and evaluating it at $t = 1$, we obtain
\begin{equation}\label{a4}
\mathcal{F}(1) = \beta \int_{0}^{\infty} ds\; sP(\beta s; \eta) = \beta^{-1} \int_{0}^{\infty} dz\; zP(z; \eta) = \beta^{-1} \mathcal{F}(\eta),
\end{equation}
from which it follows that $\langle s \rangle_t = \mathcal{F}(1)\,t^{-1/\gamma}$.
The time dependence of $\langle s \rangle_t$ coincides with that in Eq.~\eqref{a2}, the variable $t$ can be eliminated in favor of $\langle s \rangle_t$, leading to the dynamical scaling law expressed in Eq.~\eqref{dsl}.
This discussion here does not rely on any specific form of $q(r)$.

\section{Derivation of Equation~\eqref{eq2.4.0}}\label{app2}

In this appendix, we derive the IDE (Eq.~\eqref{eq2.4.0}) from Eq.~\eqref{eq2.1}, assuming the functional form $T(s) = t_{0}\,(s/s_{0})^{-\gamma}$, where $t_{0}$ and $s_{0}$ denote the characteristic time and size scales, respectively.
We begin by nondimensionalizing Eq.~\eqref{eq2.1} in the form that includes Eq.~\eqref{eq2.2}, which leads to
\begin{equation}\label{b1}
\frac{\partial \tilde{P}(\tilde{s};\tilde{t})}{\partial \tilde{t}} = -\tilde{s}^{\gamma}\tilde{P}(\tilde{s};\tilde{t})+\int_{0}^{1}dr\; \frac{q(r)}{r}  \left(\frac{\tilde{s}}{r}\right)^{\gamma} \tilde{P}\left(\frac{\tilde{s}}{r};\tilde{t}\right),
\end{equation}
where the dimensionless quantities are defined as $\tilde{s}=s/s_{0}$, $\tilde{t}=t/t_{0}$, and $\tilde{P}(\tilde{s};\tilde{t})=s_{0}P(s;t)$.
Considering the functional form of the time evolution of the mean fragment size derived in Appendix~\ref{app1}, we express $\langle s \rangle_t$ as $\langle s \rangle_t = (1/a)\, s_{0}(t/t_{0})^{-1/\gamma}$, where $a$ is a dimensionless constant.
From here, we apply a change of variables to Eq.~\eqref{b1}, using $x = s / \langle s \rangle_t$.
By defining a variable-transformed probability density function $\mathcal{P}\left(x;\tilde{t}\right)\coloneqq\tilde{P}\left(\tilde{s};\tilde{t}\right) d\tilde{s}/dx$, Eq.~\eqref{b1} is transformed into
\begin{equation}\label{b2}
\begin{aligned}
\tilde{t}\frac{\partial \mathcal{P}\left(x;\tilde{t}\right)}{\partial \tilde{t}} &+ \frac{1}{\gamma}\frac{\partial }{\partial x}\left[x\mathcal{P}\left(x;\tilde{t}\right)\right]\\
& = -\left(\frac{x}{a}\right)^{\gamma}\mathcal{P}\left(x;\tilde{t}\right)+\int_{0}^{1}dr\,\frac{q(r)}{r^{1-\gamma}}\left(\frac{x}{a}\right)^{\gamma}\mathcal{P}\left(\frac{x}{r};\tilde{t}\right).
\end{aligned}
\end{equation}
Owing to the dynamical scaling law, the time-derivative term is dropped from Eq.~\eqref{b2} in a long-time limit; then, we obtain Eq.~\eqref{eq2.4.0} by replacing $\lim_{\tilde{t}\to\infty}\mathcal{P}\left(x;\tilde{t}\right)$ with $p\left(x\right)$.

\section{Asymptotic behaviors of $p(x)$}\label{app3}

In this appendix, we derive the asymptotic behaviors of $p(x)$ as $x \to 0$ and $x \to \infty$.
We consider an IDE with $I$ given by Eq.~\eqref{eq2.4.a}.
As $x \to 0$, the second term on the right-hand side of Eq.~\eqref{eq2.4.a} becomes negligible compared with the first.
In addition, by expanding $q$ inside the integral and letting the lower limit of the integration tend to zero, the IDE can be rewritten as
\begin{equation}\label{c1}
-\frac{1}{\gamma}\frac{d}{dx}(x p(x)) + Ax^{\alpha-1}
 = 0,
\end{equation}
where the constant $A$ is given by
\begin{equation}\label{c1.0}
A =\frac{1}{\text{Beta}(\alpha,\alpha)}\int_{0}^{\infty}\frac{dw}{w^{2-\alpha}}\,\left(\frac{w}{a}\right)^{\gamma}p\left(w\right),
\end{equation}
which does not depend on $x$.
This leads to $p(x) \propto x^{\alpha - 1}$ as $x \to 0$.
On the other hand, as $x \to \infty$, the integral in Eq.~\eqref{eq2.4.a} vanishes.
In addition, the term $-(1/\gamma) p(x)$ obtained by expanding the first term becomes negligible compared to the second term.
Consequently, the IDE can be rewritten as
\begin{equation}\label{c2}
-\frac{1}{\gamma} x \frac{d}{dx} p(x) - \left( \frac{x}{a} \right)^{\gamma} p(x) = 0,
\end{equation}
which leads to $p(x) \propto \exp\left[ - (x / a)^{\gamma} \right]$ as $x \to \infty$.
Equation~\eqref{2.8} is a function that naturally connects these asymptotic behaviors.


\begin{thebibliography}{10}


\bibitem{Shorlin2000}
K.~A. Shorlin, J.~R. de~Bruyn, M.~Graham, and S.~W. Morris, 
Phys. Rev. E {\bfseries 61},  6950 (2000). 
\href{https://link.aps.org/doi/10.1103/PhysRevE.61.6950}{10.1103/PhysRevE.61.6950}

\bibitem{Yuse1993}
A.~Yuse and M.~Sano, Nature {\bfseries 362}, 329 (1993). 
\href{https://doi.org/10.1038/362329a0}{10.1038/362329a0}

\bibitem{Giorgiutti2016}
F.~Giorgiutti-Dauphin\'e and L.~Pauchard, J. Appl. Phys. {\bfseries 120}, 065107 (2016). 
\href{https://doi.org/10.1063/1.4960438}{10.1063/1.4960438}

\bibitem{Goehring2015}
L.~Goehring, A.~Nakahara, T.~Dutta, S.~Kitsunezaki, and S.~Tarafdar, 
{\em Desiccation cracks and their patterns: Formation and Modelling in Science and Nature} (John Wiley \& Sons, 2015).
\href{https://onlinelibrary.wiley.com/doi/book/10.1002/9783527671922}{10.1002/9783527671922}




\bibitem{Bacchin2018}
P.~Bacchin, D.~Brutin, A.~Davaille, E.~Di~Giuseppe, X.~D. Chen, I.~Gergianakis, F.~Giorgiutti-Dauphin{\'e}, L.~Goehring, Y.~Hallez, R.~Heyd, R.~Jeantet, C.~Le~Floch-Fou{\'e}r{\'e}, M.~Meireles, E.~Mittelstaedt, C.~Nicloux, L.~Pauchard, and M.-L. Saboungi, Eur. Phys. J. E {\bfseries 41}, 94 (2018).
\href{https://doi.org/10.1140/epje/i2018-11712-x}{10.1140/epje/i2018-11712-x}

\bibitem{Groisman1994}
A.~Groisman and E.~Kaplan, Europhys. Lett. {\bfseries 25}, 415 (1994).
\href{https://dx.doi.org/10.1209/0295-5075/25/6/004}{10.1209/0295-5075/25/6/004}

\bibitem{Allain1995}
C.~Allain and L.~Limat, Phys. Rev. Lett. {\bfseries 74}, 2981 (1995).
\href{https://link.aps.org/doi/10.1103/PhysRevLett.74.2981}{10.1103/PhysRevLett.74.2981}

\bibitem{Lecocq2002}
N.~Lecocq and N.~Vandewalle, Eur. Phys. J. E {\bfseries 8}, 445 (2002).
\href{https://doi.org/10.1140/epje/i2002-10040-2}{10.1140/epje/i2002-10040-2}






\bibitem{Akiba2017}
Y.~Akiba, J.~Magome, H.~Kobayashi, and H.~Shima, Phys. Rev. E {\bfseries 96}, 023003 (2017).
\href{https://link.aps.org/doi/10.1103/PhysRevE.96.023003}{10.1103/PhysRevE.96.023003}

\bibitem{Lilin2024}
P.~Lilin, M.~Ibrahim, and I.~Bischofberger, Sci. Adv. {\bfseries 10}, eadp3746 (2024).
\href{https://www.science.org/doi/abs/10.1126/sciadv.adp3746}{10.1126/sciadv.adp3746}

\bibitem{Ito2014a}
S.~Ito and S.~Yukawa, Phys. Rev. E {\bfseries 90}, 042909 (2014).
\href{https://link.aps.org/doi/10.1103/PhysRevE.90.042909}{10.1103/PhysRevE.90.042909}

\bibitem{Ito2014b}
S.~Ito and S.~Yukawa, J. Phys. Soc. Jpn. {\bfseries 83}, 124005 (2014).
\href{https://doi.org/10.7566/JPSJ.83.124005}{10.7566/JPSJ.83.124005}






\bibitem{Halasz2017}
Z.~Hal\'asz, A.~Nakahara, S.~Kitsunezaki, and F.~Kun, Phys. Rev. E {\bfseries 96}, 033006 (2017).
\href{https://link.aps.org/doi/10.1103/PhysRevE.96.033006}{10.1103/PhysRevE.96.033006}

\bibitem{Ito2020}
S.~Ito, A.~Nakahara, and S.~Yukawa, arXiv:2009.13691.
\href{https://arxiv.org/abs/2009.13691}{10.48550/arXiv.2009.13691}

\bibitem{Raissi2017}
M.~Raissi, P.~Perdikaris, and G.~Karniadakis, J. Comput. Phys. {\bfseries 378}, 686 (2019).
\href{https://doi.org/10.1016/j.jcp.2018.10.045}{10.1016/j.jcp.2018.10.045}

\bibitem{Karniadakis2021}
G.~E. Karniadakis, I.~G. Kevrekidis, L.~Lu, P.~Perdikaris, S.~Wang, and L.~Yang, Nat. Rev. Phys. {\bfseries 3}, 422 (2021).
\href{https://doi.org/10.1038/s42254-021-00314-5}{10.1038/s42254-021-00314-5}

\bibitem{Kovachki2023}
N.~Kovachki, Z.~Li, B.~Liu, K.~Azizzadenesheli, K.~Bhattacharya, A.~Stuart, and A.~Anandkumar, J. Mach. Learn. Res. {\bfseries 24}, 1 (2023).
\href{http://jmlr.org/papers/v24/21-1524.html}{http://jmlr.org/papers/v24/21-1524.html}

\bibitem{Ito2015}
S.~Ito, Dr. Thesis, Department of Earth and Space Science, Osaka University (2015).
\href{https://hdl.handle.net/11094/52296}{10.18910/52296}



\bibitem{Dosovitskiy2019}
A.~Dosovitskiy and J.~Djolonga, Proc. 7th Int. Conf. on Learning Representations, 2020.
\href{https://openreview.net/forum?id=HyxY6JHKwr}{https://openreview.net/forum?id=HyxY6JHKwr}

\bibitem{Bae2022}
J.~Bae, M.~R. Zhang, M.~Ruan, E.~Wang, S.~Hasegawa, J.~Ba, and R.~Grosse, Proc. 11th Int. Conf. on Learning Representations, 2023.
\href{https://openreview.net/forum?id=OJ8aSjCaMNK}{https://openreview.net/forum?id=OJ8aSjCaMNK}


\bibitem{Tancik2020}
M.~Tancik, P.~Srinivasan, B.~Mildenhall, S.~Fridovich-Keil, N.~Raghavan, U.~Singhal, R.~Ramamoorthi, J.~Barron, and R.~Ng, Adv. Neural Inf. Process. Syst., 2020, {\bfseries 33}, p. 7537--7547.
\href{https://dl.acm.org/doi/abs/10.5555/3495724.3496356}{10.5555/3495724.3496356}






\bibitem{Ramachandran2017}
P.~Ramachandran, B.~Zoph, and Q.~V. Le., arXiv:1710.05941.
\href{https://doi.org/10.48550/arXiv.1710.05941}{10.48550/arXiv.1710.05941}


\bibitem{Wang2022}
S.~Wang, H.~Wang, J.~H. Seidman, and P.~Perdikaris., arXiv:2210.01274.
\href{https://doi.org/10.48550/arXiv.2210.01274}{10.48550/arXiv.2210.01274}


\bibitem{Supplemental}
(Supplemental Materials) The architecture-dependency study and the details of the Levenberg--Marquardt algorithm are provided online.



\bibitem{Takahasi1974}
H.~Takahasi and M.~Mori, Publ. Res. Inst. Math. Sci. {\bfseries 9}, 721 (1974).
\href{https://doi.org/10.2977/prims/1195192451}{10.2977/prims/1195192451}











\bibitem{Itogithub2025}
\href{https://github.com/itoshin110/Fragment-Size-Density-Estimator-for-Shrinkage-Induced-Fracture-Based-on-PINN}{Fragment-Size-Density-Estimator-for-Shrinkage-Induced-Fracture-Based-on-PINN}

\bibitem{Nocedal1980}
J.~Nocedal, Math. Comput. {\bfseries 35}, 773 (1980).
\href{https://doi.org/10.1090/S0025-5718-1980-0572855-7}{10.1090/S0025-5718-1980-0572855-7}


\bibitem{Krishnapriyan2021}
A.~S. Krishnapriyan, A.~Gholami, S.~Zhe, R.~M. Kirby, and M.~W. Mahoney, Adv. Neural Inf. Process. Syst., 2021, {\bfseries 34}, p. 26548--26560.
\href{https://dl.acm.org/doi/10.5555/3540261.3542294}{10.5555/3540261.3542294}

\bibitem{lu2021}
L.~Lu, X.~Meng, Z.~Mao, and G.~E. Karniadakis, SIAM rev., {\bfseries 63}, 208 (2021).
\href{https://doi.org/10.1137/19M1274067}{10.1137/19M1274067}



\bibitem{LEVENBERG1944}
K.~Levenberg, Q. Appl. Math. {\bfseries 2}, 164 (1944).
\href{https://doi.org/10.1090/qam/10666}{10.1090/qam/10666}

\bibitem{Marquardt1963}
D.~W. Marquardt, SIAM J. Appl. Math. {\bfseries 11}, 431 (1963).
\href{https://doi.org/10.1137/0111030}{10.1137/0111030}

\bibitem{Jnini2025}
A.~Jnini and F.~Vella, arXiv:2505.21404.
\href{https://doi.org/10.48550/arXiv.2505.21404}{10.48550/arXiv.2505.21404}

\bibitem{Betancourt2017}
M.~Betancourt, arXiv:1701.02434.
\href{https://doi.org/10.48550/arXiv.1701.02434}{10.48550/arXiv.1701.02434}

\bibitem{Corbella2022}
A.~Corbella, S.~E.~F. Spencer, and G.~O. Roberts, Stat. Comput. {\bfseries 32}, 107 (2022).
\href{https://doi.org/10.1007/s11222-022-10142-x}{10.1007/s11222-022-10142-x}




\end{thebibliography}
\end{document}